\documentstyle[prl,aps,twocolumn,psfig]{revtex}

\tighten

\begin{document}

\draft
\preprint{to appear in Physical Review B}

\twocolumn[\hsize\textwidth\columnwidth\hsize\csname @twocolumnfalse\endcsname

\title{Resonances of dynamical checkerboard states in Josephson arrays
with self-inductance}
\author{M. Barahona$^{1}$, E. Tr\'{\i}as$^{1}$, T.P. Orlando$^{1}$,
A.E. Duwel$^{1}$, 
H.S.J. van der Zant$^{2}$, S. Watanabe$^{3}$,
and S.H. Strogatz$^{4}$}
\address{ $^{1}$Department of Electrical Engineering and Computer Science,
Massachusetts Institute of Technology, Cambridge, MA 02139\\
$^{2}$Department of Applied Physics, Delft University of Technology,
Lorentzweg 1, 2628 CJ Delft, The Netherlands\\
$^{3}$CATS, Niels Bohr Institute, Blegdamsvej 17, DK-2100, Copenhagen \O,
Denmark\\
$^{4}$Kimball Hall, Theoretical and Applied Mechanics, Cornell University,
Ithaca, NY 14853}
\date{submitted on September 27, 1996; to appear in Physical Review B}
\maketitle

\begin{abstract}

We study the dynamics of fully frustrated, underdamped Josephson arrays.
Experiments reveal remarkable similarities among the dc current-voltage
characteristics of several kinds of square and triangular arrays, where two
resonant voltages are observed.
Simulations indicate that a dynamical checkerboard solution underlies
these similarities.  By assuming such a solution, we reduce the governing
equations to three coupled pendulum equations, and thereby calculate the
voltages of the intrinsic resonances analytically.

\end{abstract}

\vskip2pc]

Discrete arrays of nonlinear oscillators can exhibit diverse spatiotemporal
patterns.  Examples include kinks in chains of coupled pendula,
neuromuscular waves in the intestine, and modulated waves and chaos in
networks of phase-locked loops~\cite{WinfreeAfraimovich}.  Although such
oscillator arrays are difficult to analyze completely, one can often use
the symmetries of the system to construct simple patterns composed of
spatially repeated ``unit cells''.  Then the governing equations reduce to a
much smaller set of equations for each unit cell.  This strategy has been
used recently to construct rotating spiral waves in a model of discrete
excitable media~\cite{Paullet}.

In this paper we present experiments, simulations, and analysis on a
broad class of discrete arrays of nonlinear Josephson-junction oscillators.
Networks ranging from single
square and triangular plaquettes to one- and two-dimensional arrays are
considered.  In each case, the junctions are identical and underdamped, and
the arrays are driven by a dc bias current.   These arrays are known to
have complicated dynamics, and only a few analytical results have been
obtained~\cite{Sohn,Octavio,2Danalytical,landsberg}.  However, one class
of arrays is relatively tractable: fully frustrated arrays,
i.e., arrays subjected to an applied magnetic field of $f=1/2$ flux quanta
per plaquette on average.  Then the ground state is highly symmetric --
the junction phases adopt a checkerboard pattern~\cite{teitel}.  This
pattern can be robust: even if the array is driven by an applied dc
current~\cite{Octavio}, or a combined ac+dc current, as in studies of giant
Shapiro steps~\cite{Sohn,Lee}, the checkerboard can persist, though now in
the form of a propagating state.

All of the fully frustrated arrays we have tested experimentally exhibit
strikingly similar dc current-voltage ($IV$) characteristics, with two
resonant voltages $V_+ \sim \Phi_0 (L_sC)^{-1/2}$ and $V_- \sim
\Phi_0(L_JC)^{-1/2}$. (Here $\Phi_0$ is the flux quantum, $L_s$ is the loop
inductance, $C$ is the junction capacitance, and $L_J =\Phi_0/2\pi I_c$ is
the Josephson inductance for a junction with a critical current $I_c$).
We show below that these similarities in the $IV$ curves can be explained
by assuming that the junction phases are organized into a dynamical
checkerboard solution.  This ansatz reduces the governing circuit
equations to three coupled pendulum equations, which in turn allows us to
obtain analytical predictions for the observed resonant voltages.

\begin{figure}[tbp]
\centerline{\psfig{file=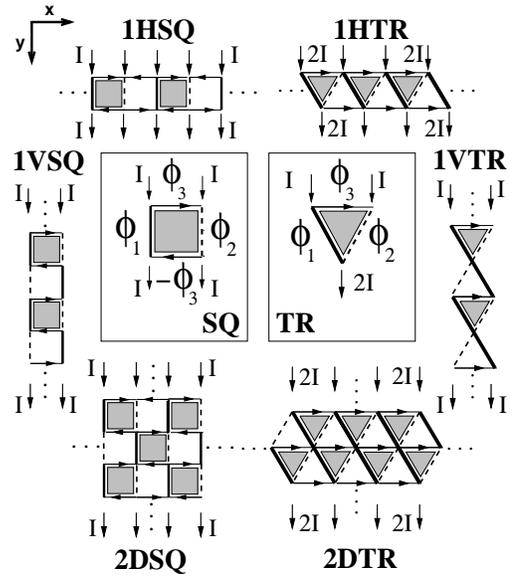,height=3in}}
\vspace{3mm}
\caption{
Arrays and their checkerboard states: Single square
(SQ) and single triangular (TR) plaquettes; 1D square (1HSQ and 1VSQ) and
1D triangular (1HTR and 1VTR) arrays; 2D square (2DSQ) and 2D
triangular (2DTR) arrays. The phases of the vertical junctions
$\phi_1$ and $\phi_2$ and the horizontal junction $\phi_3$ are
defined in the SQ and TR diagrams. The checkerboard solutions are constructed
by combining these single plaquettes with their symmetric counterparts.
\label{array_fig}}
\end{figure}

Previous authors have used the ansatz of a dynamical checkerboard,
or some other symmetric pattern, to obtain reduced equations for
frustrated Josephson arrays~\cite{Sohn,Octavio,2Danalytical,landsberg,yukon}.
Our analysis extends this work in three respects:  (1)
Inductance effects are essential to explain some of the experimental results.
Therefore, we include self-inductance in the analysis and simulations as a
first approximation. (2) The governing equations for the dynamical checkerboard
are shown to have the same algebraic form for all eight arrays in
Figure~\ref{array_fig}, up to topological factors that encode the array
geometry.   This shared structure allows a unified analysis of the
different networks. (3) Intrinsic resonant voltages are predicted analytically
from the reduced equations.

Figure~\ref{array_fig} depicts the arrays and their dynamical checkerboard
states. Junctions (not shown) exist along each branch, and current is
applied at each node along the top and bottom edges. Throughout this paper,
``vertical'' refers to the direction of current injection.
We consider junctions with identical critical currents $I_c$,
resistances $R_n$ and capacitances $C$.
Experiments have been performed on arrays of Nb-Al$_2$O$_x$-Nb
junctions~\cite{Hypres} with small damping $ \Gamma = \beta_c ^{-1/2}$,
where $5<\beta_c<25$ is the McCumber parameter.  The magnetic field is applied
normal to the arrays by an external solenoid. The two-dimensional penetration
depth $\lambda =L_J/L_s $, which can be viewed as a measure of the
discreteness of the arrays, ranges from $0.5 < \lambda <2$.
By use of a diagnostic technique~\cite{Herreapl94}, we determine
the parameters for each array: first, the measurement of  $R_n$ and
$I_c(T)$, with $I_c(0) R_n = 1.9$ mV, yields the temperature-dependent
Josephson inductance $L_J(T)$; second, from the Fiske modes of the
diagnostic we obtain $C$, $L_s$, and the mutual
inductance for nearest neighbors, which we find to be
small~\cite{Herreapl94,Bock}.  To facilitate comparisons between
different types of arrays, the triangular arrays
were built by decimating every other horizontal junction from the
corresponding square array; thus their self-inductances are equal.

\begin{figure}[t]
\centerline{\psfig{file=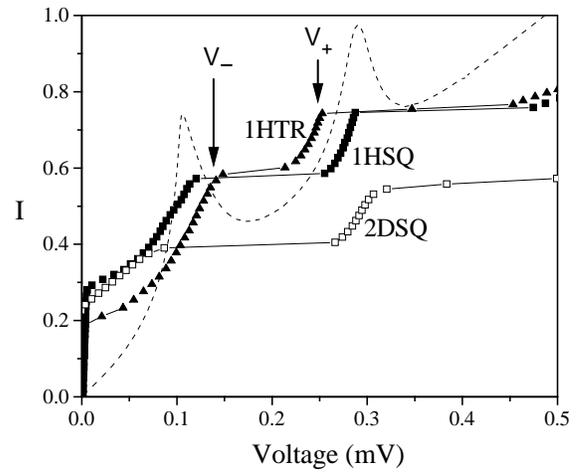,height=2.8in}}
\vspace{3mm}
\caption{
Experimental $IV$ curves for three arrays:
1HTR array ($1\times 9$ plaquettes) with $\beta_c=8$ and $\lambda=0.64$;
1HSQ array ($1\times 7$) with $\beta_c=11$ and $\lambda=0.76$;
and 2DSQ array ($7\times 7$) with $\beta_c=20$ and $\lambda=0.92$.
Dashed line, $IV$ from
harmonic balance for the 1HSQ array with the same $\beta_c$ and
an effective $\lambda_{\rm eff}=0.61$ which accounts for mutual inductance
effects. $V_+$ and $V_-$ are indicated for 1HTR.
\label{ivdata_fig}}
\end{figure}

Figure~\ref{ivdata_fig} shows the measured $IV$ curves
($I$ is the current per vertical
junction normalized by $I_c$, and $V$ is the voltage per row)
for three different geometries at $f=1/2$.
The signature of all these $IV$s is the appearance of
jumps at two resonant voltages, $V_+$ and $V_-$.
The upper step, which ends at $V_+$, is nearly vertical and independent
of temperature.
For 1HSQ and 2DSQ arrays (the arrays we have studied most
extensively), $V_+$ is also independent of the number of cells in the
$x$-direction~\cite{ET_thesis}. These results suggest
that local geometrical properties, i.e. $L_s$, determine the voltage.
A temperature-independent voltage  $V_+ \approx  \Phi_0/\sqrt{L_sC}$
is expected from dimensional arguments and
is found to be approximately correct experimentally and in simulations.
When varying the magnetic frustration, $V_+$ has the usual periodicity in
$f$ and reflection symmetry about $f=1/2$ but it is found to be almost
independent of $f$ for $0.2<f<0.5$. Thus,
although the value of $V_+$ is similar to the Eck peak in
1D parallel arrays~\cite{Herrejap94}, it does not have
the sine-like $f$--dependence observed in that case~\cite{ET_thesis}.
Moreover, this upper step does not appear for $f<0.2$ or for
$\lambda < 0.5$ as the system switches directly into row-switched states.
The lower voltage $V_-$, on the
other hand, is temperature-dependent, suggesting a dependence
on the Josephson inductance; namely,
$V_- \approx \Phi_0/\sqrt{L_J(T)C}$, up to a factor of order unity.

The dynamical origin of these two resonances is revealed through
numerical simulation of the complete arrays.
The governing equations result from current conservation at each node
and from fluxoid quantization around each loop. The current $I_j$ through each
junction, in the resistively and capacitively shunted junction model, is
 $I_j=\sin{\phi_j} + \Gamma \dot{\phi_j} +
\ddot{\phi_j}$ where the currents are in units of $I_c$, and
time is in units of $(L_J C)^{1/2}$, the inverse of the plasma
frequency. Fluxoid quantization demands that
$\sum_j \phi_j = - 2\pi f -I_{\rm cir}/\lambda$, where the sum is around
a loop, $I_{\rm cir}$ is the loop current, and only self-(loop)-inductances
are considered.  Simulated $IV$s for square arrays are consistent with
the experiments~\cite{thegang}.

Furthermore, the simulations at $f=1/2$ suggest that solutions
with wavelength equal to two plaquettes, as shown in Figure~\ref{array_fig},
underlie the observed numerics. These dynamical checkerboard solutions
are constituted by single plaquettes (SQ and TR as shown) paired with their
symmetric counterparts.
Under these symmetry constraints, the number of
relevant variables is greatly reduced and the dynamics of the
whole array is governed by a set of three coupled equations.
Moreover, the equations for each of the eight arrays in Figure~\ref{array_fig}
can be recast in the same algebraic form, up to constant factors that encode
the array topology.  In this unified formulation, the fluxoid quantization
condition at $f=1/2$ is
  \begin{equation}
      \phi_1 -\phi_2 - h \phi_3= -\pi + I_3/(\lambda m_y)
    \label{eq:fluxquant}
  \end{equation}
and the two current conservation conditions are
    \begin{equation}
         I_1 + I_2 = 2I
\hspace{.2in} {\rm and} \hspace{.2in}    I_2-I_1 = 2 m_x I_3 \; /m_y.
    \label{eq:horcurr}
   \end{equation}
Here $h$ is the number of horizontal junctions per cell
($h=2$ for square and $h=1$ for triangular arrays), and
$m_y = 1+\nu_y/2$ where $\nu_y$ is
the number of neighboring cells in the $y$-direction; likewise,
$m_x=1+\nu_x/2$.
For example, the 1HSQ array has $h=2$, $m_y=1$, and $m_x=2$,
and the 1VTR array has $h=1$, $m_y=2$, and $m_x=1$.

It is convenient to introduce three new variables:
$\phi=(\phi_1 +\phi_2)/2$, the average of the vertical phases;
$\theta= (\phi_2 -\phi_1 -\pi)/2$, which measures how much
the difference of the vertical phases differs from $\pi$;
and $\varphi = \phi_3$,  the top horizontal phase.
When the arrays have horizontal junctions, i.e. $h \neq 0$,
Eqs.~(\ref{eq:fluxquant})--(\ref{eq:horcurr}) can be
rewritten as a system of three coupled nonlinear pendulum equations:
   \begin{eqnarray}
  \label{phi_eq}
      \ddot{\phi} +\Gamma \dot{\phi} -\sin{\theta}\,\sin{\phi} &=& I \\
  \label{theta_eq}
      \ddot{\theta}+\Gamma \dot{\theta} +\cos{\phi}\, \cos{\theta} &=&
      -\lambda m_x \, (2 \theta + h \varphi)\\
  \label{varphi_eq}
       \ddot{\varphi}+ \Gamma \dot{\varphi} + \sin{\varphi} &=&
        -\lambda m_y\, (2 \theta + h \varphi).
   \end{eqnarray}

Numerical simulations indicate that, for the parameter regimes considered,
$\phi$ is approximately a uniformly whirling phase with frequency $\omega$,
whereas both $\theta$ and $\varphi$ librate sinusoidally at $\omega$.
Thus, we approximate the solution as:
$\phi = \omega t + k$, $ \theta = B \sin{(\omega t + \alpha)}$,
and $ \varphi = A \sin{\omega t}$,
where $\alpha$,  $k$, $A$, $B$ and $\omega$ are constants to be
determined for each set of driving conditions \mbox{\{$I, \Gamma, \lambda$\}}.
A harmonic balance calculation with this assumed solution
yields an $IV$ characteristic with two resonant peaks, as shown by the
dashed line in Figure~\ref{ivdata_fig} for the 1HSQ device.
These peaks are typically associated with a loss of stability of the
underlying dynamical state,
thus explaining the experimental steps.
This semi-analytic $IV$ approximates the voltages of the
resonant steps reasonably well, although it does not provide a quantitative
fit to the measured currents.  The details of this calculation will be
given elsewhere \cite{thegang}.

Approximate formulas for the resonant voltages can be obtained via a
further simplification.  Since the libration amplitudes $A$ and $B$ are
observed to be small for a wide range of bias current, we regard both
$\varphi$ and $\theta$ as small oscillations driven by the whirling
mode $\phi$. Then Eqs.~(\ref{theta_eq})--(\ref{varphi_eq})
can be linearized and,
in the limit of small damping ($\Gamma \ll 1$),
these two equations have two resonance frequencies at
   \begin{equation}
   \label{eq:omegapm}
   \omega_\pm^{2} =1/2 +\lambda \,  m_x \sigma \pm
      \left [ \left(1/2 +  \lambda \, m_x  \sigma \right)^{2} -
2 \lambda m_x \right ]^{1/2}
   \end{equation}
where $\sigma = 1+ h m_y/(2 m_x)$.
These frequencies agree with the location of the
peaks of the $IV$ from the full harmonic balance calculation.
To clarify the physical meaning of the two resonances $\omega_\pm$, consider
the limit of small inductance, $\lambda \gg 1$. Then,
$\omega_+ \approx \sqrt{2 \lambda  m_x \, \sigma} = b_+ \,\sqrt{\lambda}$
and $ \omega_- \approx  \sigma ^{-1/2} = b_- $.
The corresponding voltages are $V_+ \approx
b_+ \, \Phi_0/(2\pi \sqrt{L_s C})$
and $V_- \approx b_-\,\Phi_0/(2\pi\sqrt{L_J(T) C})$,
as guessed earlier from dimensional arguments. Thus, $V_+$ is
temperature-independent
while $V_-$ depends on $T$; and $b_+$ and $b_-$ are combinations of
topological factors of order unity.
(For the single triangle (TR), $b_+= \sqrt{3}$, as suggested by Yukon and
Lin\cite{yukon}).

\begin{figure}[t]
\centerline{\psfig{file=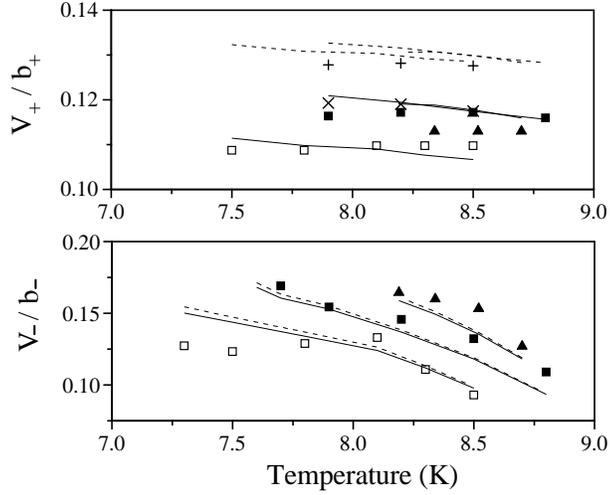,height=2.9in}}
\vspace{3mm}
\caption{
Normalized resonant voltages vs.\ $T$, measured for the
arrays presented in Fig.~\protect{\ref{ivdata_fig}}:
1HTR (solid triangles), 1HSQ (solid squares), and 2DSQ (open
squares). Dashed lines,  Eq.~(\protect{\ref{eq:omegapm}}).
Solid lines, Eq.~(\protect{\ref{eq:omegapm}})
with $L_{s,\rm eff}$ as defined in the text. Values of $V_+/b_+$
from numerical simulations of the 1HSQ array with only self-inductance
(pluses) and the full inductance matrix (crosses) are also shown.
The 1HTR array was built from a 1HSQ by decimation of every other
horizontal junction. Therefore, the geometrical unit cells are identical
and  $L_s =23$ pH and $C=300$ fF for the three arrays.
\label{data_fig}}
\end{figure}

Figure~\ref{data_fig} shows good agreement between the temperature dependence
of the experimental (squares and triangles) and predicted (lines) normalized
voltages $V_+/b_+$ and $V_-/b_-$ for 1HSQ, 2DSQ and 1HTR arrays.
The fact that Eq.~(\ref{eq:omegapm}) (dashed line)
consistently overestimates the value for $V_+/b_+$ can
be attributed to an underestimation of the inductance
since simulations with self-inductance alone (pluses in the figure)
agree well with the theory.
When simulations with the full inductance matrix are performed (crosses),
the resonant voltage moves down toward the experimental data.
The main effect of the mutual inductive coupling can be approximated
by an effective inductance $L_{s,\rm eff}=L_s \, \{1+M(\nu_x + \nu_y)\}$
where $M \approx 0.12$ is the ratio of the magnitude of the nearest-neighbor
inductance to the self-inductance in square arrays \cite{Bock}.
The solid line corresponds to  $V_+/b_+$ calculated from
Eq.~(\ref{eq:omegapm}) with this effective inductance and predicts well both
the full inductance simulations and the experimental data.
In addition, the values and temperature dependence of $V_-/b_-$ agree well with
the theory and simulations and, as expected, no inductance effects are visible.

Both the horizontal junctions and the inductance play essential roles in
the phenomena described above; without them, one or both of the observed
resonances would be lost.  To see this, consider three well-studied
limiting cases. First, if the horizontal junctions are absent, the SQ and
TR arrays are simply two-junction SQUIDS  with
inductances, and 1HSQ and 1HTR become 1D parallel arrays, which can be
regarded as discrete versions of long Josephson junctions. The dynamics of
these systems are then governed by only two equations, Eqs.~(\ref{phi_eq})
and~(\ref{theta_eq}) with $h=0$, since $\varphi$ is no longer a valid
dynamical variable. This system has a single linear resonance, at
a frequency $\omega = \sqrt{ 2 \lambda  m_x} = \omega_+ |_{h=0}$.
Second, in the limit $\lambda \rightarrow \infty$
where inductances are neglected, there is no divergence in the system
(\ref{phi_eq})--(\ref{varphi_eq}) since
$\theta \equiv - h \varphi /2$. Thus, Eqs.~(\ref{theta_eq})
and~(\ref{varphi_eq}) are combined to
eliminate $\lambda \, (2 \theta + h \varphi)$ and the dynamics of the system
is then governed by only two equations: Eq.~(\ref{phi_eq}) and a
reduced equation for $\theta$,
\begin{equation}
\ddot{\theta} + \Gamma \dot{\theta} +
\frac{1}{\sigma} \frac{h}{2}
\left \{ \sin \left ( \frac{2}{h} \theta \right)
+\frac{m_y}{m_x} \cos \theta \, \cos \phi \right \} = 0.
\end{equation}
(These equations include as a special case the 2DSQ system previously studied
in~\cite{Sohn,Octavio}. Note also that when $\lambda \rightarrow \infty$,
the SQ and 2DSQ equations are identical but this is no longer true when
inductances are taken into account). In conclusion, in the absence of
inductance there is a single linear resonant frequency
$\omega = \sigma ^ {-1/2} = \omega_- |_{\lambda \rightarrow \infty}$.
Third, if there are neither horizontal junctions nor inductances, the sole
dynamical equation~(\ref{phi_eq}) has no resonances.

These limiting cases also suggest a heuristic explanation for the origin of
the resonances.   At $V_-$, as the checkerboard slides across the array, it
strongly excites spin waves, or equivalently, ``ringing'' oscillations of
the junctions at an eigenfrequency close to their characteristic plasma
frequency $(L_J(T) C)^{-1/2}$. This resonance requires horizontal junctions
but not inductance, and is temperature-dependent.  In contrast, the resonance
at $V_+$ is due to excitation of electromagnetic modes of the array.  This
resonance requires inductance but not horizontal junctions, and depends
only on local geometrical properties of the array that are
temperature-independent.  The corresponding eigenmode is mainly related to
oscillations in the induced flux per cell.

The role of the
horizontal junctions can be further explored by considering
an anisotropic network where horizontal and vertical junctions are fabricated
with the same process and differ only in their area. Then the ratio of their
critical currents, $\eta = I_{c1}/I_{c3}$, parametrizes the anisotropy of the
array. In this case, the results from
equations~(\ref{phi_eq})-(\ref{eq:omegapm}) are still valid with a
renormalized $m_y' = m_y \, \eta$, which specifies the participation
of the horizontal junctions in the dynamics of the array.

Finally, we emphasize that our analysis is based on the assumption of a
dynamical checkerboard state.  Unfortunately, very little is known about
the conditions for its global stability.  If it is unstable, or if the array
organizes itself in some alternative stable
state, the dynamics are not yet understood.  For instance, simulations of
2DTR arrays with low $\beta_c$  seem to show a different solution
(the ``ribbon state") where the horizontal
junctions are essentially inactive and a checkerboard is formed by double
cells~\cite{yukon}. This solution is analogous to the striped columnar
dynamical configuration observed in $f=1/2 \,$ Shapiro steps in square
arrays~\cite{Dominguez}.
In both cases, the arrays effectively behave as a collection of in-phase
rows and, thus, have only one resonant voltage at $V_+|_{h=0}$.
Experiments on triangular arrays will address these issues
separately~\cite{triang}.
Also, if the junctions are highly underdamped, the checkerboard state in 2DSQ
arrays can slide chaotically~\cite{Octavio}.  The conditions for the
stability and the temporal periodicity of the checkerboard state, and the
dynamics associated with other possible states, are challenging problems
for future investigation.

We thank P.\ Caputo, S.\ Yukon, N.\ Lin and A.\ Ustinov for useful discussions
and for sharing their unpublished results on triangular arrays.
Research supported in part by NSF grants DMR-9402020 and
DMS-9500948 and by Rome Laboratory (AFMC) grant \mbox{F 30602-96-2-0059}.
AED and ET acknowledge support from the NSF Graduate Fellowship Program.

\end{document}